# Features of interaction of microwaves with He II

V.D. Khodusov and A.S. Naumovets

*V.N.Karazin Kharkov National University, 4 Svobody sq., 61077 Kharkov, Ukraine*

*Based on interrelation between the thermodynamic and electromechanical phenomena in superfluid helium, the explanation of experimentally found features of microwave interaction in the frequency range of 40-200 GHz is given. Due to fast roton-roton and roton-phonon interactions resonant excitation on frequency correspond to minimal roton energy relaxes and forms a wide pedestal. Alongside these fast processes, there are also slower processes of rotons' scattering by microwave photons taking place, which lead to additional absorption of energy of resonant microwaves and to the appearance of a narrow resonant peak on the background of a wide pedestal. The theoretical explanation of the influence which streams exert on resonant absorption of microwaves is given. The critical velocity of stream at which absorption of microwaves was replaced by their radiation is found.

## 1. INTRODUCTION

Several studies[1,2] showed that, during the distribution of second sound waves and torsion vibrations in helium, a number of special displays of electric properties of He II, which specify the link between the thermodynamic and electromechanical phenomena, have been found. Research on absorption of the microwaves in a frequency range of 40-200 GHz in superfluid helium, has shown number of unexpected results: resonant absorption of the microwaves on frequencies corresponding to energy of rotons[3-6] was revealed, meanwhile the narrow line of absorption near to the minimal energy of rotons was observed, and the carried measurements of temperature dependence of this narrow line corresponded to change of the minimal energy of rotons. It was also found that, there is an influence of velocity of relative movement of normal $\mathbf{v}_n$ and superfluid $\mathbf{v}_s$ components on the character of resonant interaction of the microwaves. With the increase in velocity of relative movement, the factor of absorption of the microwaves decreased, at the excess of some critical velocity absorption of waves was replaced by their radiation. Similar effects take place in the plasma physics at the description of linear and nonlinear Landau attenuation of plasma waves on particles, and also excitation of these waves by streams of particles, moving with the velocity higher than some critical[7].

One study[8] has revealed that in strong electric fields, a splitting of a narrow line of microwaves absorption occurs on the two lines symmetrically located relative to the line in absence of an electric field. Thus the value of splitting increases with growth of an electric field practically linearly that corresponds to linear Stark effect.

These experimental facts testify to strengthening displays of electric properties in helium at temperatures below $T_\lambda$. Although, they specify a special role, which can play quasi-particles (phonons, rotons) for an explanation of these experiments. Attempts of a theoretical explanation of these features have been made in works[9-14]. In the given work the way of an explanation of features of interaction of the microwaves with superfluid HeII, based on the account of influence of electric fields on quasi-particles and their kinetics is offered. In the temperature range in which the experiments were carried out (1,4 K – 2,3 K), the basic role is played by rotons which define both thermodynamic and kinetic properties of He II. Due to rapid roton-roton and roton-phonon interactions, the hydrodynamic regime is set in gas of quasi-particles. The explanation of these effects can be lead, if one considers the quasi-particle energy to be linear on an electric field[15].

## 2. RAMAN SCATTERING OF ELECTROMAGNETIC WAVES FROM QUASI-PARTICLES IN HeII

Interaction of electromagnetic waves with superfluid helium has previously been studied, both theoretically and experimentally at the description Raman scattering of light from phonons, second sound and rotons[16-21]. In Refs. 17 and 18 a two-roton scattering of light was experimentally observed. As the roton momentum considerably exceeds (on some orders) the photon momentum (visible light and microwaves) for monitoring of Raman scattering of electromagnetic waves from rotons, as follows from momentum conservation, in scattering process should take part two rotons[19]. Energy and momentum conservation permit a born of two rotons by photon and photons scattering from rotons.

For the first process energy and momentum conservation laws are:
$$\hbar\omega_1 = \hbar\omega_2 + \varepsilon_3 + \varepsilon_4; \qquad \hbar\mathbf{k}_1 = \hbar\mathbf{k}_2 + \mathbf{p}_3 + \mathbf{p}_4, \qquad (1)$$
where $\omega_{1,2}$ and $\mathbf{k}_{1,2}$ – frequencies and wave a vectors of the incident and reflected electromagnetic wave accordingly, $\varepsilon_{3,4}$ and $\mathbf{p}_{3,4}$ – energies and momentums of rotons respectively. The scattering in this case will involve two excitations with opposite momenta $\mathbf{p}_3 \approx -\mathbf{p}_4$.

Taking into account that the roton energy is $\varepsilon_r = \Delta + \frac{(p-p_0)^2}{2\mu}$, from (1) will obtained for Stokes (red) satellite:

$$\hbar(\omega_1 - \omega_2) \geq 2\Delta + \frac{(p_3 - p_0)^2}{2\mu}, \qquad (2)$$

where $p_0$ – value of momentum at which energy or roton has a minimum equal to $\Delta$, $\mu$ – effective weight of roton.

The density of number of energy states of rotons per unit of volume will be equal:

$$\rho(\varepsilon) = \frac{\sqrt{2\mu}}{4\pi^2 \hbar^3} \frac{\left(p_0 + \sqrt{2\mu(\varepsilon - \Delta)}\right)^2}{\sqrt{\varepsilon - \Delta}}. \qquad (3)$$

From Eq. (3) it follows, that the basic role in these processes will play rotons with minimal energy $\varepsilon \approx \Delta$. Then, from Eq. (2) follows, that realizes two-roton light absorption that was observed in Refs. 17 and 18. Theoretical explanation of these experiments and also finding matrix elements of photon and roton interactions are given in Refs. 19 and 20.

For the second process describing photons scattering from rotons, conservation laws of energy and momentum are:

$$\hbar\omega_1 + \varepsilon_3 = \hbar\omega_2 + \varepsilon_4; \qquad \hbar\mathbf{k}_1 + \mathbf{p}_3 = \hbar\mathbf{k}_2 + \mathbf{p}_4. \qquad (4)$$

In contrast to the first case, here we have $\mathbf{p}_3 \approx \mathbf{p}_4$.

From the compatibility condition of system (4) follows the next condition imposed on frequencies of the incident and reflected electromagnetic waves, for Stokes satellite:

$$0 \leq (\omega_1 - \omega_2) \leq \frac{2(p_3 - p_0)|\mathbf{k}_1 - \mathbf{k}_2| + \hbar|\mathbf{k}_1 - \mathbf{k}_2|^2}{2\mu}. \qquad (5)$$

This condition, differs from (2), in that, it shows that frequencies $\omega_1$ and $\omega_2$ differ from each other a little. Indeed, in the conditions of experiments[17], carried estimations give the following result for a case $p_3 \approx p_0$, $\mathbf{k}_1 \approx -\mathbf{k}_2$, $k_1 \approx 10^5 cm^{-1}$: $(\omega_1 - \omega_2) \approx 2 \cdot 10^6 s^{-1}$. For the first case, under the same conditions, if one was to take the value for the roton minimal energy at temperature $T = 1.4K$ $\Delta = 8.65K$, from Eq. (2) we would receive $(\omega_1 - \omega_2) \approx 2.27 \cdot 10^{12} s^{-1}$. Intensity of Stokes line of Raman scattering of light on rotons due to the second process following on from the research[20], will be $e^{-\frac{\Delta}{T}}$ time less, compared to scattering caused by the first process. In addition it will be in other frequency ranges (MHz). We consider, that despite the small intensity of this line, it can be experimentally revealed by modern equipment.

A special interest in studying Raman scattering, represents methods of resonant combinational scattering of light[22]. Thus instances, when frequency of light coincides with own frequencies of elementary excitations are possible. In this case, there will be an imposing of two effects: a forced resonant excitation of quasi-particles by electromagnetic wave and photon scattering from quasi-particles. If a frequency of an electromagnetic wave reaches roton energy, as follows from energy conservation, the processes of two roton born will be forbidden.

## 3. RESONANCE INTERACTION OF MICROWAVES WITH SUPERFLUID HELIUM

It is possible to explain the resonant excitation of rotons by electromagnetic fields of microwaves on frequency $\omega_{res} \approx \Delta/\hbar$ in means of hydrodynamic approximation, which is realized in HeII due to rapid roton-roton and roton-phonon interactions, at the temperature range where experiments were carried out.

In Ref. 23, based on the fact that the length of the thermal de Broglie wave is greater than atomic dimensions, the nonlocal relation between density and pressure was obtained:

$$\rho(\mathbf{r}) = \int h(\mathbf{r} - \mathbf{r}_1) P(\mathbf{r}_1) d\mathbf{r} . \qquad (6)$$

This nonlocality is essential in describing the short-wave (roton) excitations. In this case, was found the following relation between the Fourier components $\rho_k$ and $P_k$:

$$\rho_k = \frac{\hbar^2 k^2}{\varepsilon^2(k)} P_k . \qquad (7)$$

where $\varepsilon(k)$ – spectrum quazi-particles.

Free oscillations of the medium described by the equation:

$$\frac{\partial^2 \rho'}{\partial t^2} - \Delta P' = 0 . \qquad (8)$$

The equation of state in general terms has the appearance:

$$P = P(\rho, T, \mathbf{E}, \mathbf{w}) .$$

If the driving force acting as external electric field $\mathbf{E} = \mathbf{E}_0 e^{i\omega t}$, then the equation for forced oscillations with accounting of main terms and quazi-particles attenuation $\gamma$ will take on form:

$$\frac{\partial^2 P_k}{\partial t^2} + \gamma \frac{\partial P_k}{\partial t} + \frac{\varepsilon^2(k)}{\hbar^2} P_k = -\frac{\partial \rho}{\partial \mathbf{E}} \frac{\partial^2 \mathbf{E}}{\partial t^2} \frac{\varepsilon^2(k)}{\hbar^2 k^2}. \qquad (9)$$

It is well-known[24] that, in the case of the forced oscillation, the distribution of intensity in a resonant line of electromagnetic waves absorption, normalized to unit is:

$$dI = \frac{\gamma}{\left(\omega - \omega_{res}\right)^2 + \gamma^2/4} d\omega, \qquad (10)$$

where $\omega_{res} = \varepsilon(k)/\hbar$. Factor of quasi-particle attenuation $\gamma$ define a half-width of a resonant curve.

In the case of rotons the equation for forced oscillations take on form (9), where $\varepsilon = \varepsilon_r$ and $\gamma = \gamma_r$. Roton-roton and roton-phonon interactions define finite life time of rotons, which account leads to degradation of a resonant line and formation of a plateau, which is observed in experiments in a vicinity of resonant frequency[6, 8]. Indeed, received half-width of a resonant line on frequency $\omega_{res} = 1,1 \cdot 10^{12} Hz$ in Ref. 6 at $T = 1.4K$ it is equal $\gamma_r \approx 10^{11} s^{-1}$, that corresponds by the order of magnitude to inverse roton lifetime $\tau_r^{-1}$ at the same temperature[25, 26].

Notice, if the resonant frequency will correspond to the maxon energy $\omega_{res} = 1,9 \cdot 10^{12} Hz$ then one can to determine life time of maxon by measuring of half-width of resonant curve.

### 4. THE NARROW RESONANT LINE. INFLUENCE OF THE STREAMS.

We consider, that a sharp resonant line, which is observed on a background of a pedestal (Rayleigh wings), is caused by the slower process of scattering of photons from rotons. At resonance condition by the time $1/\gamma_r \Box 10^{-11} s$ quasi-local distribution function of rotons is established:

$$n_{r0} = \left( \exp\left[\frac{\varepsilon + \mathbf{pw}}{T}\right] - 1 \right)^{-1} \qquad (11)$$

where $\mathbf{w} = \mathbf{v_n} - \mathbf{v_s}$ – relative velocity.

The change in unit time the number of microwave photons $\Delta N_1$ with energies $\hbar \omega_1$ due to induced processes of scattering fotons from rotons can be written in the form:

$$\frac{\partial \Delta N_1}{\partial t} = \Delta N_1 \int |\Phi(1,3;2,4)|^2 N_2 (n_{04} - n_{03}) \delta(\hbar(\omega_1 - \omega_2) + \varepsilon_3 - \varepsilon_4) \times$$
$$\times \delta(\mathbf{p}_4 - \mathbf{p}_3 - \hbar(\mathbf{k}_1 - \mathbf{k}_2)) \frac{d^3 p_4 d^3 p_3}{(2\pi\hbar)^6} \frac{d^3 k_2}{(2\pi)^3} \quad (12)$$

Given that the momentum of photons is much smaller than the roton momentum, we can expand $n_{04}$ and obtain:

$$n_{04} - n_{03} = \frac{n_{03}}{T} \left( -\hbar(\omega_1 - \omega_2) + \hbar(\mathbf{k}_1 - \mathbf{k}_2)\mathbf{w} \right) \quad (13)$$

If we seek the solution this equation in the form $\Delta N \sim e^{-\gamma t}$, we will obtain

$$\gamma = -\frac{1}{T} \int |\Phi(1,3;2,4)|^2 n_{03} N_2 \left( (\omega_1 - \omega_2) - (\mathbf{k}_1 - \mathbf{k}_2)\mathbf{w} \right) \times$$
$$\times \frac{2\mu}{|\mathbf{k}_1 - \mathbf{k}_2|\sqrt{(p_3 - p_0)^2 + 2\mu\hbar(\omega_1 - \omega_2)}} \frac{2\pi p_3^2 dp_3}{(2\pi\hbar)^3} \frac{d^3 k_2}{(2\pi)^3} \quad (14)$$

Not calculating the attenuation factor, makes it possible to define these boundary frequencies, from the condition:

$$\omega_2 \leq \omega_1 \leq \omega_2 + \frac{2|p_3 - p_0||\mathbf{k}_1 - \mathbf{k}_2| + \hbar|\mathbf{k}_1 - \mathbf{k}_2|^2}{2\mu} \quad (15)$$

Considered that the frequencies lie at same range at the vicinity of resonant, it is easy to receive:

$$\frac{\Delta}{\hbar} - \frac{2|p_3 - p_0||\mathbf{k}_1 - \mathbf{k}_2| + \hbar|\mathbf{k}_1 - \mathbf{k}_2|^2}{4\mu} \leq \omega_1 \leq \frac{\Delta}{\hbar} + \frac{2|p_3 - p_0||\mathbf{k}_1 - \mathbf{k}_2| + \hbar|\mathbf{k}_1 - \mathbf{k}_2|^2}{4\mu}$$

From this relation it follows, that both resonant frequency and limiting values of the permitted frequencies of the microwaves vary with temperature basically the same way as an energy gap in a roton spectrum, which corresponds to experimental results.

The width of the base of a resonant curve, if $\mathbf{k}_1 \approx -\mathbf{k}_2$ for the thermal rotons $|p_3 - p_0| \approx \sqrt{2\mu T}$ is defined by the expression:

$$\Delta \omega \approx 2 \left( \sqrt{2\mu T} k_1 + \hbar k_1^2 \right) / \mu \quad (16)$$

If $T = 1.4\ K$, $k_1 \approx 37.7\ cm^{-1}$, then $\Delta\omega \approx 4.7 \cdot 10^5\ Hz$. It is in the order of the magnitude agreement with which was observed in experiment.

From the expression (14) for $\gamma$, shows that a critical speed $W_{cr}$ exists at which $\gamma = 0$. At speeds greater then $W_{cr}$ absorption of microwaves is replaced by their radiation. Calculating the damping factor $\gamma$ the square matrix element of the photon-roton interaction can be taken outside the integral, assuming in it rotons momentum by the order of $p_0$ and $\mathbf{k}_1 \approx \mathbf{k}_2$.

After calculating the remaining integral, we obtain the following estimate of critical velocity: $W_{cr} = \dfrac{8}{3}\dfrac{\hbar k_1}{\mu}$. Substituting the values of $k_1$ and $\mu$ obtains $W_{cr} \sim 10^{-2}$ cm / sec. By comparing the calculated factor of attenuation with the experimentally measured it is possible to obtain order estimates of the matrix element of interaction of the microwave photons and rotons.

## 5. CONCLUSION

In this paper we propose a new mechanism to explain the features of the interaction of microwaves with superfluid helium. We consider that forced resonant excitations by electric field of microwave occurs. Due to the fast roton-roton and roton-phonon interactions are excitation relaxes and forms a broad pedestal. In addition to these fast processes are much slower processes of scattering rotons on microwave photons, resulting in additional energy absorption resonance of microwaves and the formation of a narrow resonance peak on the background of a broad pedestal. It is shown that the resonance line with temperature changes as well as the minimum energy rotons. A theoretical explanation of the influence of flow on resonant absorption of microwaves is obtained.

## REFERENCES

1. A. S. Rybalko, *Low Temp. Phys*. **30**, 994 (2004)
2. A. S. Rybalko and S. P. Rubets, *Low. Temp. Phys*. **31**, 623 (2005)
3. A. S. Rybalko, S. P. Rubets, E. Ya. Rudavskii, V. A. Tikhiy, S. A. Tarapov, R. V. Golovashchenko, and V. N. Derkach, *Phys. Rev*. B **76**, 140503(R) (2007)

4. A.S. Rybalko, S.P. Rubets, E.Ya. Rudavskii, V.A. Tikhiy, R.V. Golovashchenko, V.N. Derkach, and S.I. Tarapov, *Low. Temp. Phys.* **34**, 254 (2008)
5. A.S. Rybalko, S.P. Rubets, E.Ya. Rudavskii, V.A. Tikhiy, S.A. Tarapov, R.V. Golovashchenko, and V.N. Derkach, *Low. Temp. Phys*. **34**, 497 (2008)
6. A.S. Rybalko, S.P. Rubets, E.Ya. Rudavskii, V.A. Tikhiy, U. M. Poluektov, R.V. Golovashchenko, V.N. Derkach, S.A. Tarapov, and O. V. Usatenko, *Low. Temp. Phys.* **35**, 1073 (2009)
7. A. I. Akhiezer, I. A. Akhiezer, R. V. Polovin, A. G. Sitenko, and K. N. Stepanov, *Plasma Electrodynamics*, edited by D. ter Haar, vol.1, Oxford, New York, Pergamon Press (1975)
8. A.S. Rybalko, S.P. Rubets, E.Ya. Rudavskii, V.A. Tikhiy, R. Golovashchenko, V.N. Derkach, and S.I. Tarapov, *e-print arXiv:* 0807.4810 (2008).
9. A. M. Kosevich, *Low Temp. Phys*. **31**, 37 (2005); **31**, 839 (2005).
10. V.D. Khodusov, *Visnyk KhNU*, **642**, #3/25/, 79 (2004)
11. D.M. Litvinenko, V.D. Khodusov, *Visnyk KhNU,* **721**, #1/29/, 31(2006)
12. L.A. Melnikovsky. Polarization of dielectrics by acceleration. http://www.arXiv:cond-mat/0505102(v.3, August(2006))
13. V.D. Natsik. *Low Temp. Phys.*, **31**, 915 (2005)
14. V.M.Loktev, M.D.Tomchenko, *Low Temp. Phys*., **34,** 262(2008)
15. V.D. Khodusov, A.S. Naumovets, *Visnyk KhNU,* **899,** #2/46/, 97 (2010)
16. M. Woolf, P. Platzmann, and M. Cohen, Phys. Rev. Lett. **17**, 294 (1966)
17. T. J. Greytak, and J. Yan, Phys. Rev. Lett. **22**, 987 (1969)
18. T. J. Greytak, R. Woerner, J. Yan, and R. Benjamin, Phys. Rev. Lett. **25**, 1547 (1970)
19. J. W. Halley, Phys. Rev. **181**, 338 (1969)
20. M. Stephen, Phys. Rev. **187**, 279 (1969)
21. V. L. Ginzburg, Zh. Eksp. Teor. Fiz., **13** 243 (1943)
22. R. Martin, and L. Falikov, *Topics in Applied Physics, V8, Light Scattering in Solids*, edited by M. Cardona (Springer-Verlag, Berlin, 1975), p. 101.
23. N. Adamenko, K. E. Nemchenko, and I. V. Tanatarov, Phys. Rev. B 67, 104513 (2003)
24. L. D. Landau and E M. Lifshitz, *Mechanics*, Section 26, Pergamon Press, Elmsford, N.Y. (1976)
25. L. D. Landau, I. M. Khalatnikov, JETP, 19(7), 637 (1949)
26. I. N. Adamenko, N. R. Belyaev, V. I. Tsiganok, Low. Temp. Phys, 14(9), 899 (1988)